\documentclass[a4paper,11pt]{article}
\pdfoutput=1 

\usepackage{jinstpub} 

\usepackage{lineno}
\usepackage{placeins}
\usepackage{graphicx}
\usepackage{url}
\usepackage{subcaption}

\usepackage{siunitx}

\title{First results on new helium based eco-gas mixtures for the Extreme Energy Events Project}

\author[a,b]{M. Abbrescia}
\author[c,d]{C. Avanzini}
\author[c,d]{L. Baldini}
\author[e]{R. Baldini Ferroli}
\author[c,d,l]{G. Batignani}
\author[f]{M. Battaglieri}
\author[g,h]{S. Boi}
\author[d,1]{E. Bossini\note{Corresponding author.}}
\author[i]{F. Carnesecchi}
\author[j]{F. Cavazza}
\author[h]{C. Cicalò}
\author[k,j,l]{L. Cifarelli}
\author[l]{F. Coccetti}
\author[m]{E. Coccia}
\author[n]{A. Corvaglia}
\author[o,p,l]{D. De Gruttola}
\author[o,p,l]{S. De Pasquale}
\author[q]{L. Galante}
\author[l,j]{M. Garbini}
\author[l,r]{I. Gnesi}
\author[w]{F. Gramegna}
\author[s,f]{S. Grazzi}
\author[j,i,l]{D. Hatzifotiadou}
\author[t,u,l]{P. La Rocca}
\author[v]{Z. Liu}
\author[s,u]{G. Mandaglio}
\author[j]{A. Margotti}
\author[w]{G. Maron}
\author[b]{M. N. Mazziotta}
\author[g,h]{A. Mulliri}
\author[j,l]{R. Nania}
\author[j,l]{F. Noferini}
\author[x]{F. Nozzoli}
\author[k,j]{F. Palmonari}
\author[y,n]{M. Panareo}
\author[n]{M. P. Panetta}
\author[z,d]{R. Paoletti}
\author[aa]{C. Pellegrino}
\author[f]{L. Perasso}
\author[j,l]{O. Pinazza}
\author[i]{C. Pinto}
\author[l, e]{S. Pisano}
\author[t,u,l]{F. Riggi}
\author[ab]{G. Righini}
\author[o,p,l]{C. Ripoli}
\author[b]{M. Rizzi}
\author[k,j]{G. Sartorelli}
\author[j]{E. Scapparone}
\author[ac,r]{M. Schioppa}
\author[k,j]{G. Scioli}
\author[z,d]{A. Scribano}
\author[j,l]{M. Selvi}
\author[ad,f]{M. Taiuti}
\author[d]{G. Terreni}
\author[s,u]{A. Trifirò}
\author[s,u]{M. Trimarchi}
\author[aa]{C. Vistoli}
\author[ae]{L. Votano}
\author[i,v]{M. C. S. Williams}
\author[l,k,j,i,v]{A. Zichichi}
\author[v,i]{R. Zuyeuski}

\affiliation[a]{Dipartimento di Fisica dell’Università e del Politecnico di Bari,\\Via Amendola 173, 70125 Bari, Italy}
\affiliation[b]{INFN, Sezione di Bari,\\Via Orabona 4, 70126 Bari, Italy}
\affiliation[c]{Dipartimento di Fisica,\\Università di Pisa, Largo Bruno Pontecorvo 3, 56127 Pisa, Italy}
\affiliation[d]{INFN, Sezione di Pisa,\\Largo Bruno Pontecorvo 3, 56127 Pisa, Italy}
\affiliation[e]{INFN, Laboratori Nazionali di Frascati,\\Via Enrico Fermi 54, 00044 Frascati, Italy}
\affiliation[f]{INFN, Sezione di Genova,\\Via Dodecaneso, 33, 16146 Genova, Italy}
\affiliation[g]{Dipartimento di Fisica, Università di Cagliari,\\S.P. Monserrato-Sestu, Monserrato (CA), 09042, Italy}
\affiliation[h]{INFN, Sezione di Cagliari,\\Complesso Universitario di Monserrato, S.P. per Sestu, 09042, Monserrato (CA), Italy}
\affiliation[i]{CERN,\\Esplanade des Particules 1, 1211 Geneva 23, Switzerland}
\affiliation[j]{INFN, Sezione di Bologna,\\Viale Carlo Berti Pichat 6/2, 40127 Bologna, Italy}
\affiliation[k]{Dipartimento di Fisica e Astronomia, Università di Bologna,\\Viale Carlo Berti Pichat 6/2, 40127 Bologna, Italy}
\affiliation[l]{Museo Storico della Fisica e Centro Studi e Ricerche "E. Fermi",\\Via Panisperna 89/a, 00184 Roma, Italy}
\affiliation[m]{Gran Sasso Science Institute,\\Viale Francesco Crispi 7,  67100 L'Aquila, Italy}
\affiliation[n]{INFN, Sezione di Lecce,\\Via per Arnesano. 73100, Lecce, Italy}
\affiliation[o]{Dipartimento di Fisica, Università di Salerno,\\Via Giovanni Paolo II, 132, 84084 Fisciano SA, Italy}
\affiliation[p]{INFN ,Gruppo Collegato di Salerno,\\Complesso Universitario di Monte S. Angelo ed. 6 via Cintia, 80126, Napoli, Italy}
\affiliation[q]{Teaching and Language Lab, Politecnico di Torino,\\Corso Duca degli Abruzzi 24, Torino, Italy}
\affiliation[r]{INFN, Gruppo Collegato di Cosenza,\\via Pietro Bucci, Rende (Cosenza), Italy}
\affiliation[s]{Dipartimento di Scienze Matematiche e Informatiche, Scienze Fisiche e Scienze della Terra, Università di Messina,\\Viale Ferdinando Stagno d'Alcontres 31, 98166 Messina (ME), Italy}
\affiliation[t]{Dipartimento di Fisica, Università degli Studi di Catania,\\Via. S. Sofia 64, 95123 Catania (CT), Italy}
\affiliation[u]{INFN, Sezione di Catania,\\Via. S. Sofia 64, 95123 Catania (CT), Italy}
\affiliation[v]{ICSC World laboratory,\\Geneva, Switzerland}
\affiliation[w]{INFN, Laboratori Nazionali di Legnaro,\\Viale dell'Università 2, 35020 Legnaro, Italy}
\affiliation[x]{INFN Trento Institute for Fundamental Physics and Applications,\\Via Sommarive, 14, 38123 Povo TN, Italy}
\affiliation[y]{Dipartimento di Matematica e Fisica, Università del Salento,\\Via per Arnesano. 73100, Lecce, Italy}
\affiliation[z]{Dipartimento di Scienze Fisiche, della Terra e dell’Ambiente, Università di Siena,\\Via Roma 56 - 53100 Siena}
\affiliation[aa]{INFN-CNAF,\\Viale Carlo Berti Pichat 6/2, 40127 Bologna}
\affiliation[ab]{CNR, Istituto di Fisica Applicata "Nello Carrara",\\Via Madonna del Piano 10, 50019 Sesto Fiorentino (FI), Italy}
\affiliation[ac]{Dipartimento di Fisica, Università della Calabria,\\via Pietro Bucci, Rende (Cosenza), Italy}
\affiliation[ad]{Dipartimento di Fisica, Università di Genova,\\Via Dodecaneso, 33, 16146 Genova GE}
\affiliation[ae]{INFN, Laboratori Nazionali del Gran Sasso,\\Via G. Acitelli 22, 67100 Assergi (AQ), Italy}

\emailAdd{edoardo.bossini@pi.infn.it}

\abstract{The Extreme Energy Events (EEE) Project, a joint project of the Centro Fermi (Museo Storico della Fisica e Centro Studi e Ricerche "E.Fermi") and INFN, has a dual purpose: a scientific research program on cosmic rays at ground level and an intense outreach and educational program. The project consists in a network of about 60 tracking detectors, called telescopes, mostly hosted in Italian High Schools. Each telescope is made by three Multigap Resistive Plate Chambers, operated so far with a gas mixture composed by 98\% C$_2$H$_2$F$_4$ and 2\% SF$_6$. Due to its high Global Warming Potential, a few years ago the EEE collaboration has started an extensive R\&D on alternative mixtures environmentally sustainable and compatible with the current experimental setup and operational environment. Among other gas mixtures, the one with helium and hydrofluoroolefin R1234ze gave the best result during the preliminary tests performed with two of the network telescopes. The detector has proved to reach performance levels comparable to those obtained with previous mixtures, without any modification of the hardware. We will discuss the first results obtained with the new mixture, tested with different percentages of the two components.
}


\begin{document}
\maketitle
\flushbottom

\section{Introduction} \label{sec:intro}

\begin{figure}[htbp]
\centering 
  \includegraphics[width=0.51\linewidth]{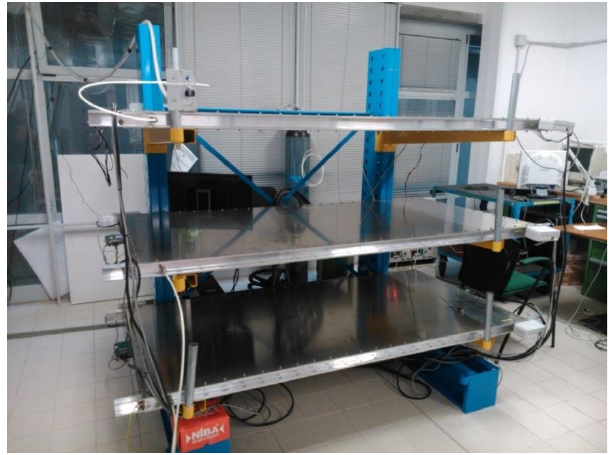}
  \includegraphics[width=0.3\linewidth]{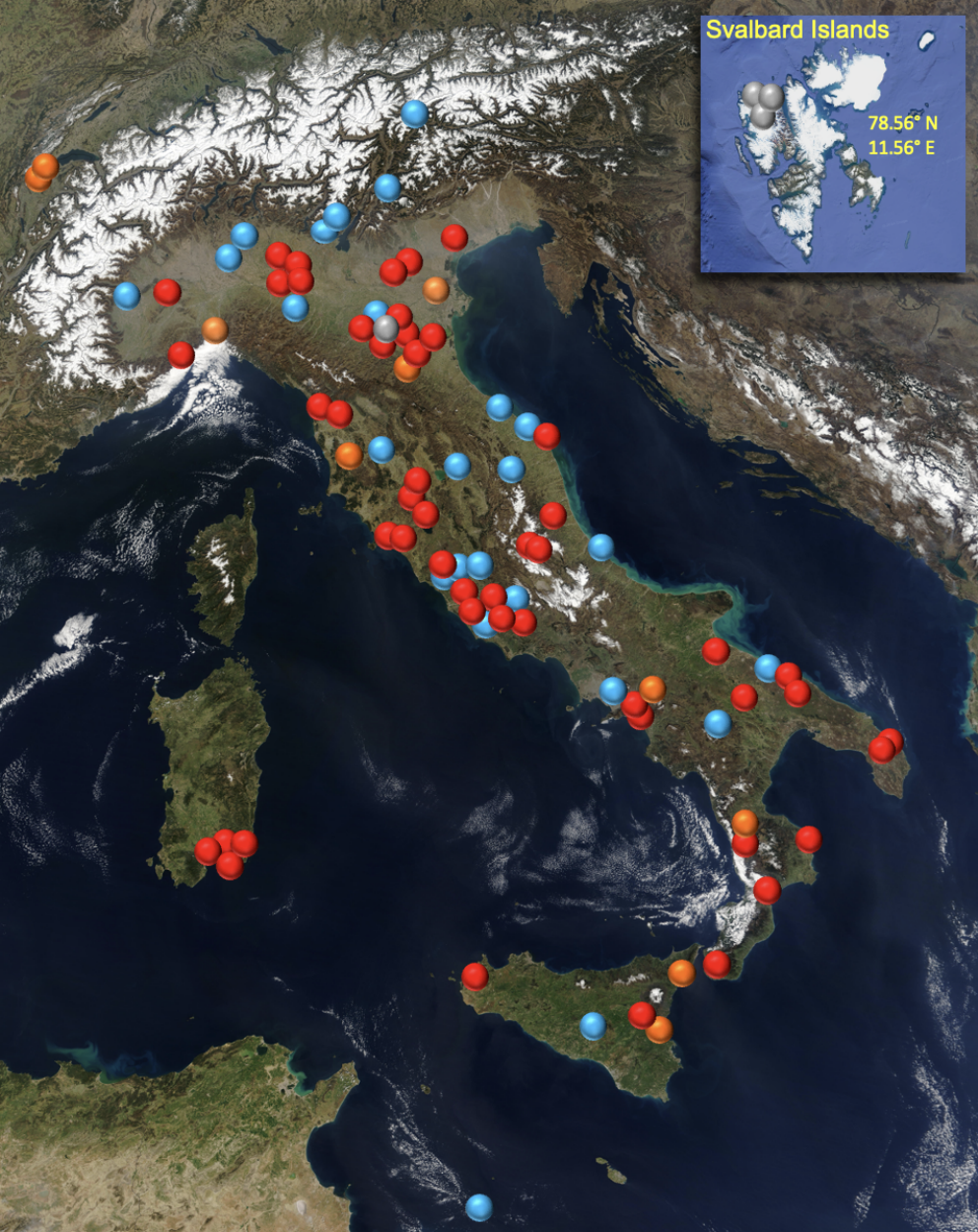}
\caption{\label{fig:overview} On the left, a picture of one of the EEE telescopes. On the right, the geographical distribution of the schools participating to the project with (red dots) or without (blue dots) a telescope. Some telescopes are installed in INFN sites or at CERN (orange dots).}
\end{figure}

The Extreme Energy Events (EEE) Project \cite{original} is an experiment based on a network of about 60 cosmic-ray measuring stations (called telescopes) installed mostly in High Schools all over Italy (Fig. \ref{fig:overview}).

Its peculiarity is that the students of the schools involved in the project have the unique opportunity to participate in the construction of the detectors at CERN, in the installation inside their own schools and in the commissioning, operations and monitoring of the telescope all over the yearly data taking periods. This makes the project unique among the various outreach programs, usually restricted to few days laboratory experience for the students.

Telescope data are centrally collected at the INFN-CNAF data center in Bologna, where the Data Quality Monitoring and data reconstruction are automatically performed. Each telescope is able to detect and track the traversing particles with multi-tracking capability and assign an \emph{absolute timestamp} to each particle using the Global Positioning System (GPS). Cosmic rays detected by individual telescopes can be thus correlated (offline) and data analyses on extensive air showers are possible.
The performance of the detectors \cite{performance} and the wide geographical distribution of the telescopes allow for a broad research program on cosmic rays at ground level. EEE has performed studies of several physics cases, among them: search for coincidences between near telescopes \cite{NearCoinc}, study of the muon flux decrease due to solar events \cite{forbush}, study of cosmic muon anisotropy at sub-TeV scale \cite{subTev}, study of muon decay into up-going events \cite{upgoing}, search for long distance correlations between extensive air showers \cite{EAS}.

\begin{figure}[htbp]
\centering 
\includegraphics[scale=0.5]{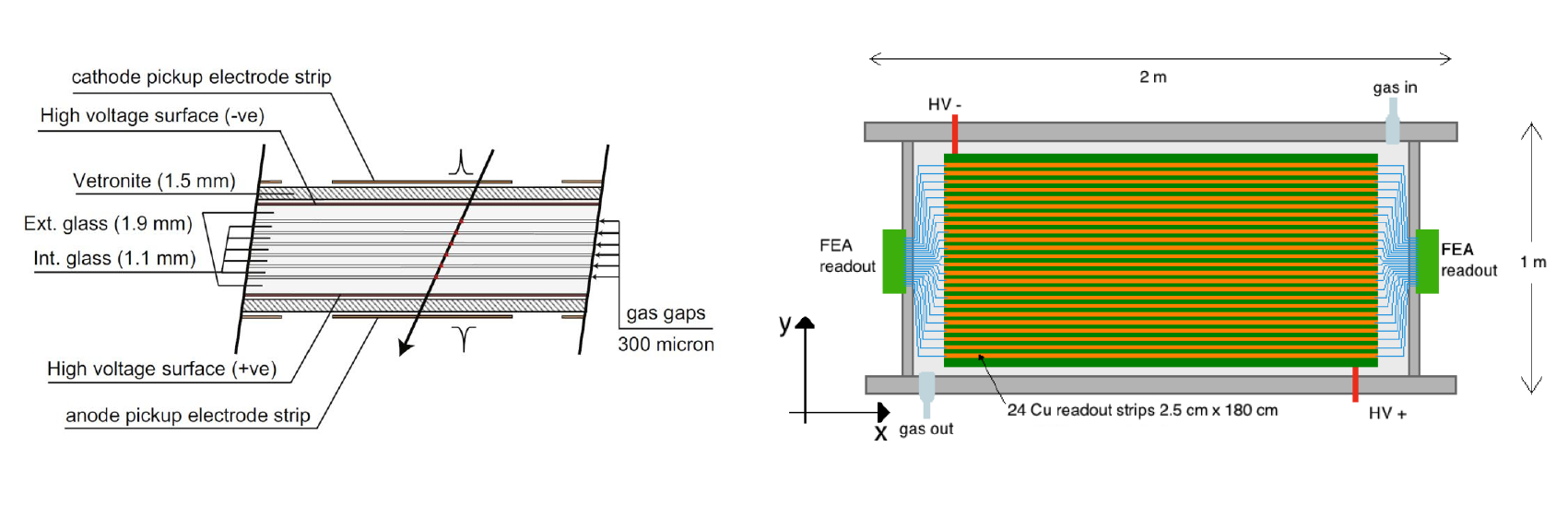}
\caption{\label{fig:detector} On the left, a schematic representation of a six gap MRPC stack. The outer glasses are thicker than the inner ones to ensure better mechanical rigidity. On the right, the schematic top view of one MRPC with the 24 top strips read out by the two front-end boards. The top strips are paired with the bottom strips, providing a differential readout scheme, and each pair is treated as a single readout channel.}
\end{figure}

The telescopes are made of 3 Multigap Resistive Plate Chambers (MRPC) separated by 50 cm, as shown in Fig. \ref{fig:overview}. 
The active volume is formed by the gaps between seven parallel and equidistant thin glasses, as shown in Fig. \ref{fig:detector} on the left. The total active surface of the detector is 158x82 cm$^2$ \cite{MRPCDetails}. In 15 years of activity, two sets of telescopes have been produced, one with a \SI{300}{\micro\meter} gap and another with a \SI{250}{\micro\meter} gap. The external surface of the two outermost glasses is coated with semiconductive paint which is connected to a voltage generator to produce a uniform electric field within the gaps. The passage of an ionizing particle through the detector generates ion-electron pairs and the latter start a multiplication process if the electric field is sufficiently high. This charge in the gaps induces a signal on the conductive strips facing both sides of the glass stack. There are 24 strips per side, parallel to each other with a 3.5 cm pitch. Fig. \ref{fig:detector} on the right shows the top view of the chamber once the aluminum lid and honeycomb panel have been removed. Each strip on one side of the glass stack is aligned to the strip on the other side and connected on both sides to the front-end electronics providing a differential readout pattern, so in the rest of the article each pair of strips is treated as a single strip reading.

Whenever an avalanche appears in the detector, an induced electric signal travels to both ends of the strips, where it is discriminated and digitized by the NINO chips \cite{NINO}, which are fast 8-channel discriminators designed with a full differential architecture, located on the front-end boards. Each front-end board, two per chamber, also generates a Local Trigger (LT) signal when at least one strip has a signal above threshold, providing a total of six LT signals for the central trigger. The digitized output of the NINO chips follows the Low-Voltage Differential Signaling (LVDS) standard, with an output signal duration, here referred as Time Over Threshold (TOT), which depends from the total input charge. The NINO is followed by a Time to Digital Converter (the CERN HPTDC \cite{HPTDC}), capable of measuring the arrival time of both the leading and trailing edges of the input signals. The HPTDC is configured to perform the time measurements on both edges with a binning of 100 ps.
It is therefore possible to acquire a precise timestamp for the time of arrival of the signal, together with the measurement of the TOT. When a coincidence of all the LTs is detected, the coincidence signal is used as a trigger to acquire the signals of all strips. The reconstruction algorithm, as described in Sec. \ref{sec:test}, can then use the time information from both strip ends to reconstruct a 2-dimensional hit on the chamber, and assign the timestamp to it. Precise timing is crucial to measure some of the particle characteristics (i.e. the velocity of the particle hitting all chambers of the telescope through the time of flight between the upper and lower chambers) and for a precise reconstruction of the impact point of the particle with each telescope chamber. The time associated to a hit in the EEE MRPCs depends, also because of (threshold) effects in the readout electronics, on the amplitude and total charge of the signals. This effect, here referred as Time Walk (TW), can be partially corrected offline using the signal TOT, enhancing the precision of the apparatus on the measure of the hit time. With He-based mixtures we have observed a broadening of the TOT distribution, a hint of a possibly broader charge spectrum which could result in a larger TW effect. Further studies are underway.

In each chamber, hits close in space and time are merged in clusters, then used to reconstruct the trajectory of the particle inside the telescope. The trajectory is assumed to be a straight line. The absolute particle timestamp is finally computed merging the particle timestamp computed in the local time reference with the synchronization signal provided by the GPS. The uncertainty on the absolute particle timestamp is usually dominated by the precision of the time precision of the GPS, of the order of few tens of ns.
\section{New eco-mixtures} \label{sec:ecogas}
Until the end of 2021 the MRPCs have been fluxed with a gas mixture of
98\% C$_2$H$_2$F$_4$ (R134a) and 2\% SF$_6$, both GreenHouse Gases(GHG), with total Global Warming Potential (GWP) $\sim$ 2030. In order to reduce the operational costs of the telescopes and the emissions of GHGs, a dedicated campaign to minimise gas leaks from the telescopes, which was started in 2019 and ended after COVID-19 shutdown, allowed to reduce the gas flux to $\sim$ 1 l/h (from the previous 2-3 l/h) for the large majority of the telescopes, corresponding to a volume exchange rate of about few per cent per hour. Notice that the addition of a gas recirculation system is also being explored to further reduce gas consumption \cite{EEE:recirc}. Currently, a gas recirculation prototype developed at CERN is under test in one of the EEE telescopes. Nevertheless, the search for new eco-friendly gas mixtures has become crucial for the EEE Project, especially given its important role in outreach and student education.
Therefore the EEE collaboration has decided to phase out the gas mixture in use and start an R\&D on alternative mixtures environmentally sustainable. Several physics experiments all over the world are pursuing the same strategy, making the search for new eco-friendly mixtures one of the most relevant topics in the field of gaseous detector development.
In the R\&D some strict requirements are posed on the typology and performance of the new gas mixture, deriving from budget constraints and from the security regulation in force in the schools where the telescopes are located:
\begin{itemize}
    \item only non flammable, non toxic gases are allowed;
    \item to match the requirements of the existing mixers, only binary mixtures can be used;
    \item the detector must be able to operate with a maximum bias voltage of 20 kV;
    \item the front-end electronics must be able to handle the new signals;
    \item the performance of the detector with the new mixture should not have any negative impact on the physics program of the experiment;
    \item the cost of the mixture should be in line with the old one.
\end{itemize}

Among all the constraints, the most limiting are represented by the restriction to binary mixtures and the upper limit on the bias voltage. Several results are indeed available on new eco-mixtures for RPC detectors, but all of them make use of three or more gases.
The strategy adopted was to replace the R134a with the hydrofluoroolefin (HFO) R1234ze (C$_3$H$_2$F$_4$), the most similar molecule with low GWP = 4 and compliant with the security requirements, and to add an almost equal percentage of helium (He) or CO$_{2}$ to the mixture, with the effect of reducing the operating voltage within the allowed range \cite{Abbrescia_2016}. A pure HFO1234ze is indeed expected to require a higher bias voltage, higher than the one which can be currently  generated. It is worth noticing that with both CO$_{2}$ and He, the total GWP remains below 10.
The expected drawback of the strategy is represented by a reduced streamer suppression capability of these two eco-friendly mixtures with respect to the standard mixture with SF$_6$. Both CO$_{2}$ and He-based compounds have been extensively tested in the EEE collaboration.
In particular, the mixture made of HFO1234ze (simply HFO in the rest of the text) and He has been tested on the telescope located in the Rende (codename REND-01) site, hosted in an INFN and University of Calabria laboratory, providing the best results to date. Note that, following some tests, we discarded HFO-CO$_{2}$ mixtures since we noticed a larger streamer contamination with respect to HFO-He mixtures, given the same efficiency.
\section{Test setup and procedure}\label{sec:test}
The results of the R\&D program reported here have been obtained testing the middle chamber fluxed with the HFO-He mixture, while operating the 2 outer chambers of the telescope with the “standard” (R134a+SF$_{6}$) mixture.

The external chambers are used as reference for trigger and tracking.
The data, collected by triggering on the coincidence of these reference chambers, have been analyzed offline with a dedicated algorithm. 
As previously discussed, chamber signals are digitized at both ends of the strips, generating End Hits (EHs). An EH contains the leading edge time and the TOT of the signal. The first step of the reconstruction is the selection and pairing of EHs. The HPTDC is set to acquire all EHs within a time match window of 500 ns, set with a proper latency with respect to the trigger arrival time. The matching window is further reduced in the offline reconstruction to $\sim$100 ns. EHs are ordered in time and for each strip end only the first EH found in the offline match window is retained. The probability of having a second EH in the offline match window is negligible.
If a strip has EHs on both ends, a hit on the chamber is formed. While the Y coordinate is directly extrapolated from the strip identification number, the longitudinal coordinate X is computed from the difference of the times of arrival of the 2 EHs, providing a 2-dimensional hit position. The average of the two arrival times, insensitive to the hit position, is in turn used to assign a precise timestamp to the hit, providing a 4D measurement (Z being fixed by the vertical position of the chamber).

Next, the clusterization is performed through an iterative procedure applied to each chamber. Firstly, the algorithm forms a hit list ordered as a function of the Y coordinate. The first hit is promoted to cluster and removed from the hit list, then the algorithm searches for another hit closer than 10 cm to the cluster. If found, it is added to the cluster and removed from the list. The search starts back from the first hit still in the list and goes ahead till the list is empty or no more hit matching the cluster is found. In case the list contains other hits, the procedure starts again, creating a new cluster. It is important to note that the distance between a cluster and a hit is the minimum distance between the hit and all the hits already assigned to the cluster. The clusterization procedure has been checked both by visual inspection of several events and by checking the compatibility of the timestamps of clusterized hits. Finally, the cluster coordinate is computed as the average of all hits coordinates. In the present analysis the information of the TOT has not been used to correct the hit timestamps. The timestamp of a cluster has been defined as the timestamp of the hit with the lowest time of arrival. This definition allows to achieve better performance compared to the average of all hit timestamps. For further improvement, work is ongoing to establish the best TW correction algorithm to be applied or to test the effect of using a weighted average of all time measurements in the cluster, with weights derived from the TOTs of the hits.
The track reconstruction algorithm, after the hit clusterization, checks if exactly one cluster is present in both triggering chambers, in practice selecting events with a single track to avoid ambiguities in the reconstruction. If the condition is met, it generates a candidate track using the clusters from the two reference chambers. The candidate track is then projected (in both space and time) in the chamber under test.
To reduce the background, dominated by spurious coincidences and upgoing particles, the following selection criteria are applied for the candidate tracks to be used in the final computation of the efficiency:
\begin{itemize}
    \item a particle speed $\beta$ in the range $0.75< \beta <1.25$ (within errors);
    \item a track projection on the chamber under test within a fiducial area, defined with a clearance of 15 cm from the edge of the active surface.
\end{itemize}

Events with tracks passing the selection criteria are then used to check the efficiency of the chamber under test. 
The chamber under test is considered efficient if a cluster is found within 15 cm and 10 ns from the extrapolated track hit. If more than one cluster is matching such condition, the closest in space is retained for the computation of the time-space residuals.

On top of the efficiency other parameters are computed as a function of the bias voltage, among which:
\begin{itemize}
    \item the streamer fraction, defined as the fraction of efficient events with a matching cluster in the test chamber made by more than 3 hits, hence called streamer (streamers are not excluded from the analysis);
    \item the average cluster size, defined as the average number of hits forming the matching cluster (even if defined as streamer);
    \item the time residual, defined as the time difference between the matching cluster time and the extrapolated track hit time;
    \item the spatial residuals, defined as the differences between the coordinates of the matching cluster center and the extrapolated impact point of the track.
\end{itemize}

The results reported in the next section have been obtained with different HFO and He relative percentages (50/50, 60/40, 70/30 and pure HFO) and compared with the "standard" mixture. Gas flow has been kept around 1 l/h. For each mixture a High Voltage (HV) scan on the chamber under test has been performed, keeping the other two chambers at a fixed HV. As anticipated in Sec. \ref{sec:ecogas}, mixtures with large fractions of HFO are expected to have a significant increase in the operating voltage, above 20 kV. To produce such a bias voltage, above the actual reach of the existing power supply units of the EEE telescopes, a different high voltage system from CAEN \cite{CAEN} has been used for the tests reported herein. The system was able to deliver up to 24 kV differential bias voltage to the chambers.
\section{Results} \label{sec:results}

\begin{figure}
\begin{subfigure}{.53\textwidth}
  \centering
  \includegraphics[width=1\linewidth]{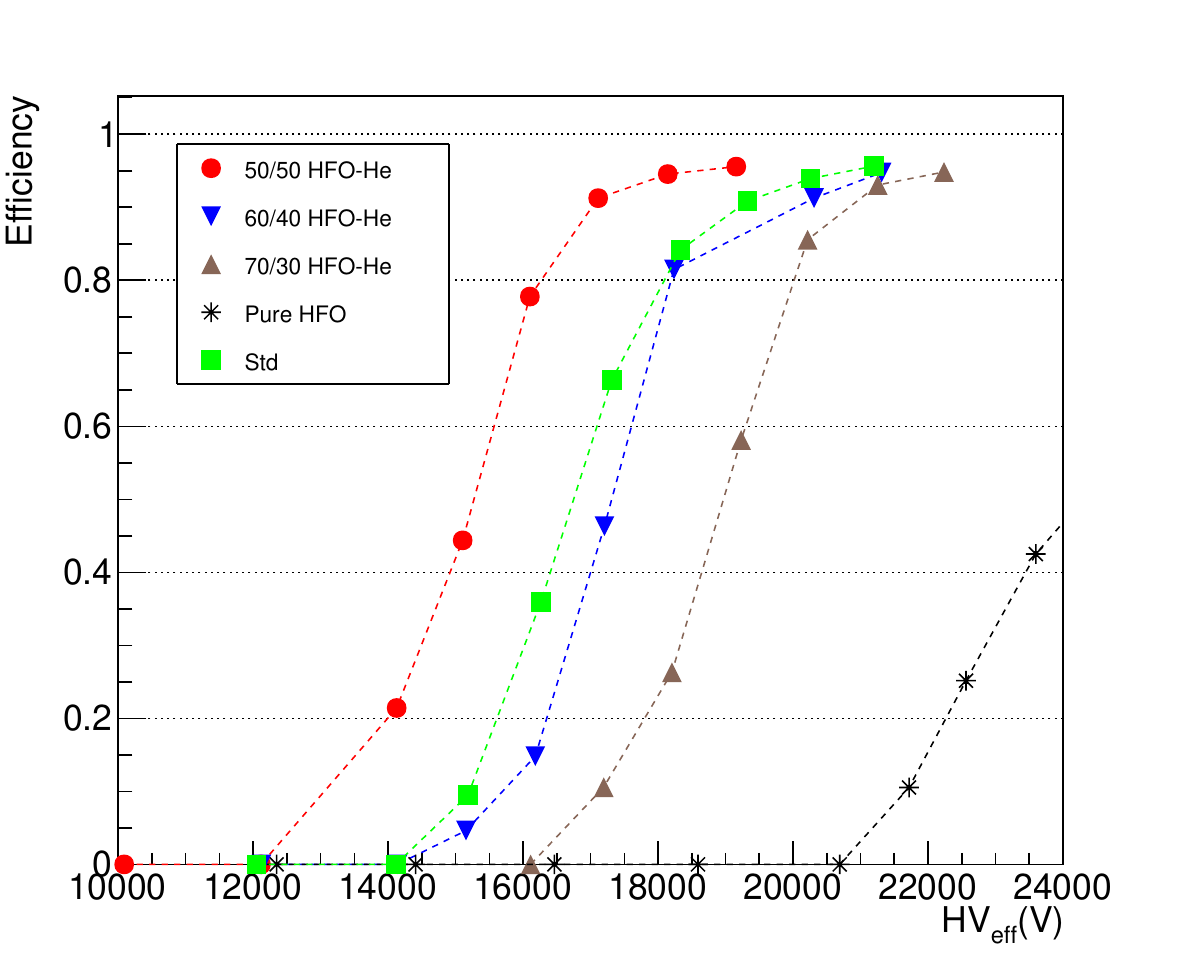}
  \caption{Efficiency scan}
  \label{fig:EfficiencyS1}
\end{subfigure}%
\begin{subfigure}{.46\textwidth}
  \centering
  \includegraphics[width=1.1\linewidth]{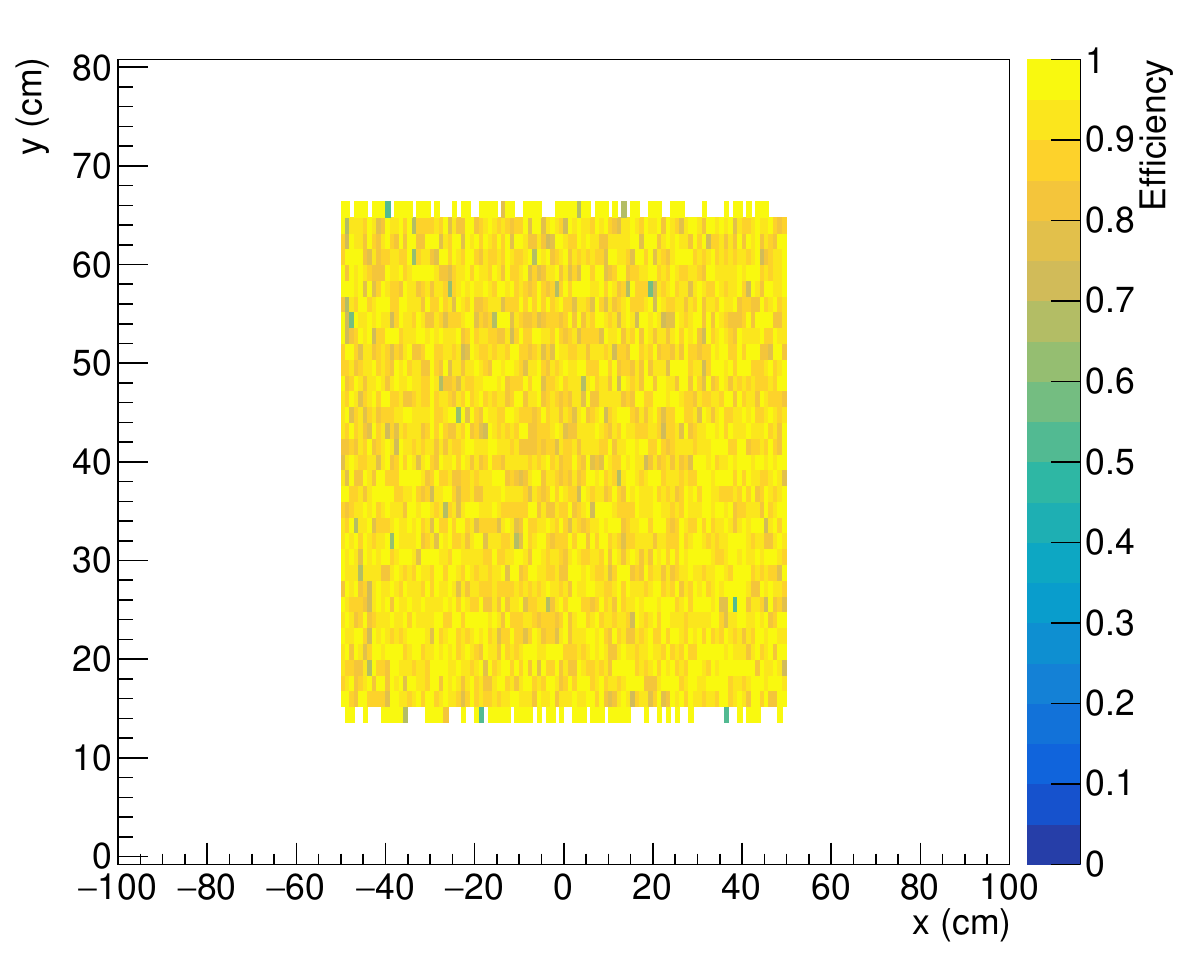}
  \caption{Efficiency map}
  \label{fig:EfficiencyS2}
\end{subfigure}
\caption{On the left, the scan of efficiency for the chamber under test with different mixtures as a function of the applied effective bias voltage. On the right, the efficiency map for the 50/50 HFO-He mixture and an effective bias voltage of $\sim18$ kV in the fiducial area.}
\label{fig:Efficiency}
\end{figure}

In Fig. \ref{fig:EfficiencyS1} the efficiency of the chamber as a function of the effective bias voltage $HV_{eff}$ using different gas mixtures is reported.
The effective bias voltage is compensated for temperature and pressure effect, according to the formula $HV_{eff}=HV*\frac{P_{ref}}{P}*\frac{T}{T_{ref}}$, where $P_{ref}=1010$ mbar and $T_{ref}=293.15$ $^\circ$K \cite{Abb1,Abb2}.
The data show, as expected, a reduction of the HV working point as the percentage of He increases. A mixture 60/40 of HFO and He respectively, provides very similar results in terms of efficiency with respect to the standard mixture.
An efficiency plateau above 90\% can be reached with a bias voltage below the 20 kV upper limit of the current experimental setup, using a mixture with at least 40\% of He. 
The uniformity of the chamber efficiency in the fiducial area can be seen in the plot of Fig. \ref{fig:EfficiencyS2}, for the 50/50 mixture and an effective bias voltage of $\sim18$ kV. The X-Y position is the one extrapolated on the test chamber using the two external reference chambers.

\begin{figure}
\begin{subfigure}{.5\textwidth}
  \centering
  \includegraphics[width=1\linewidth]{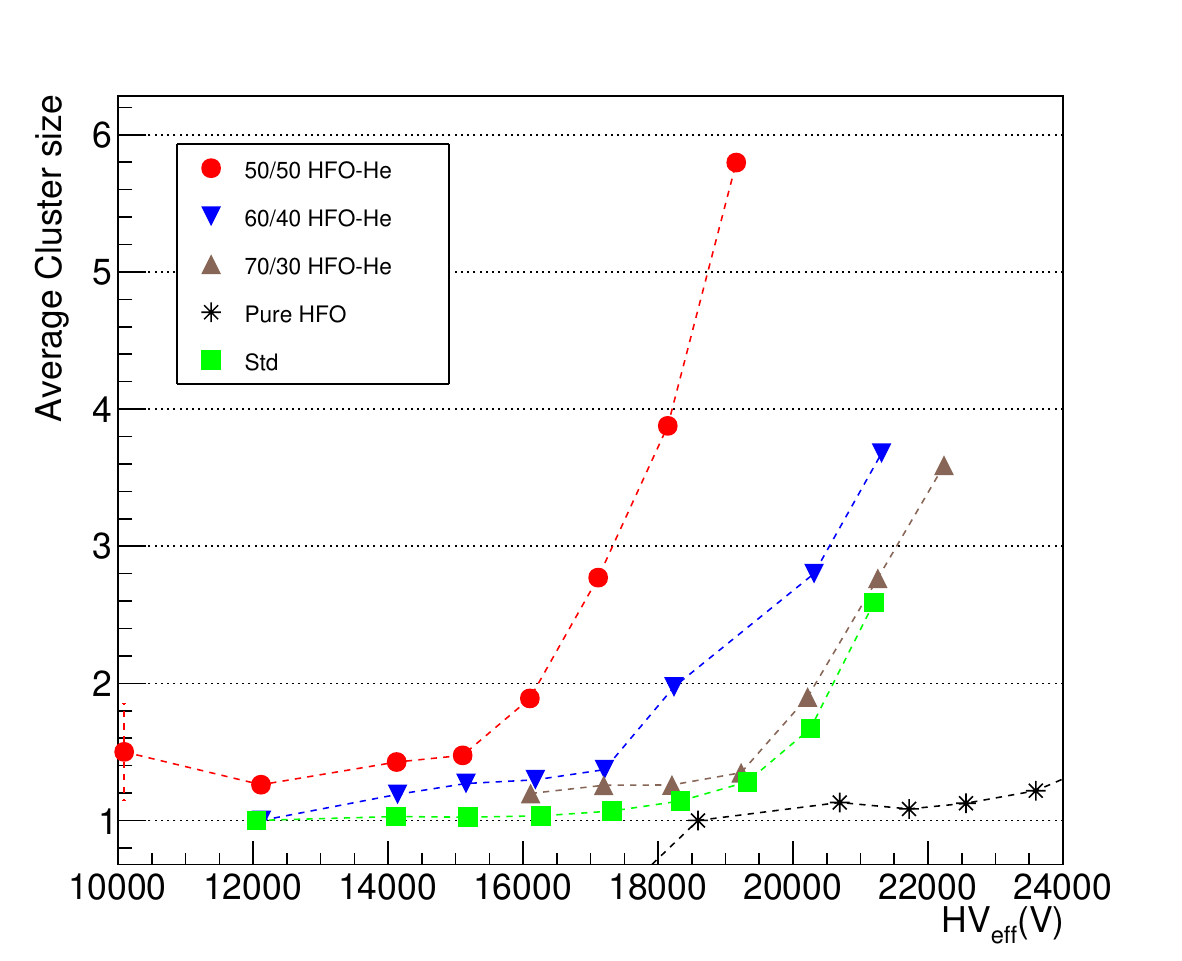}
  \caption{Cluster size Vs HV}
  \label{fig:ClusterStreamS1}
\end{subfigure}%
\begin{subfigure}{.5\textwidth}
  \centering
  \includegraphics[width=1\linewidth]{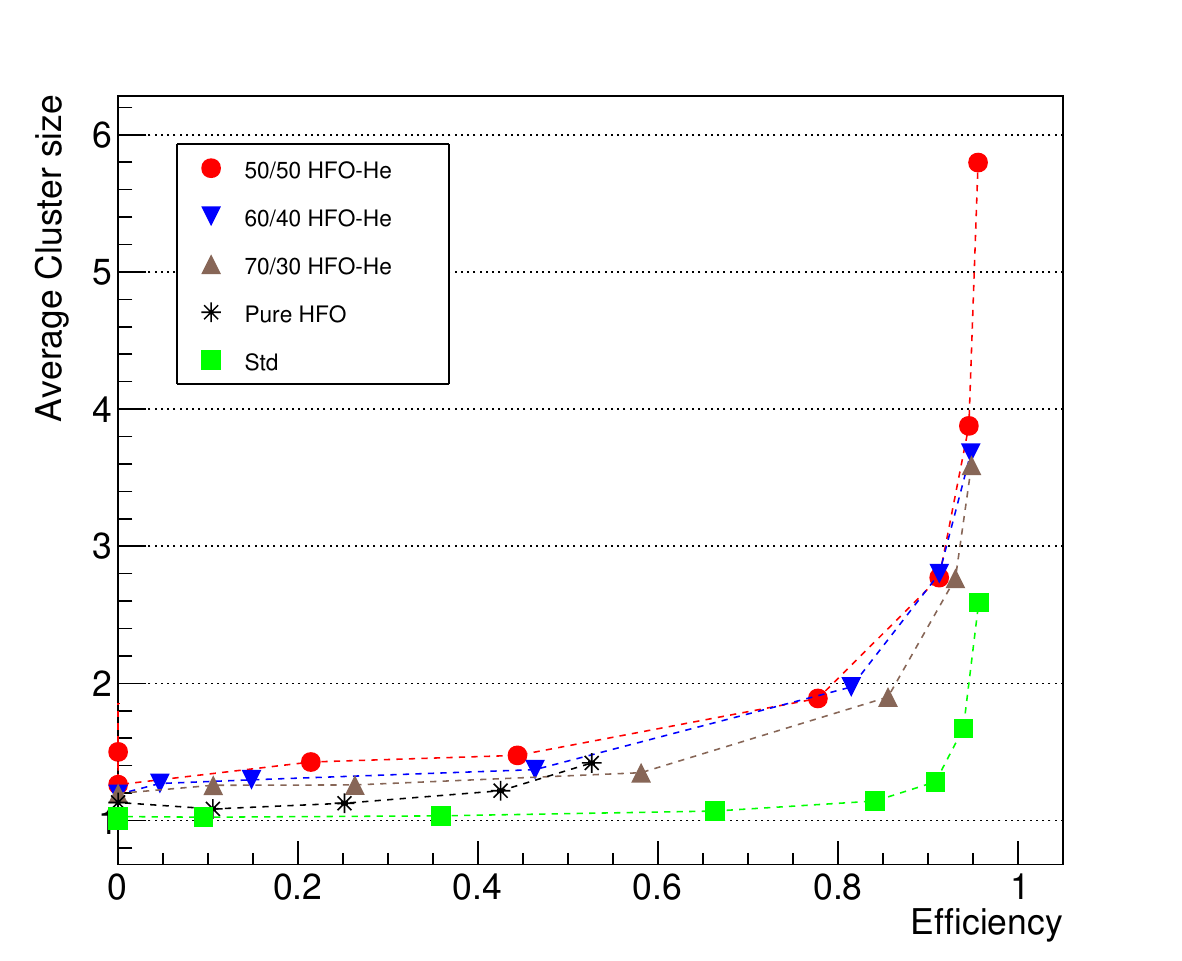}
  \caption{Cluster size Vs Efficiency}
  \label{fig:ClusterStreamS2}
\end{subfigure}
\begin{subfigure}{.5\textwidth}
  \centering
  \includegraphics[width=1\linewidth]{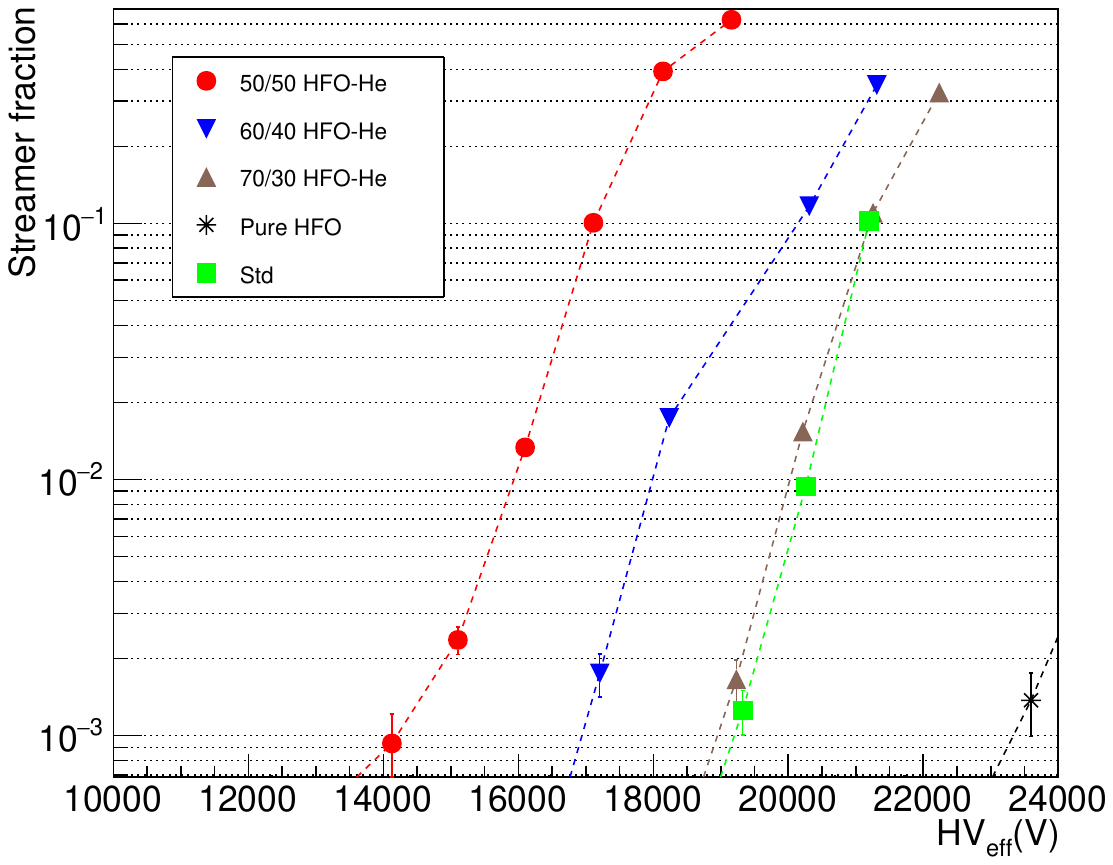}
  \caption{Streamer Vs HV}
  \label{fig:ClusterStreamS3}
\end{subfigure}
\begin{subfigure}{.5\textwidth}
  \centering
  \includegraphics[width=1\linewidth]{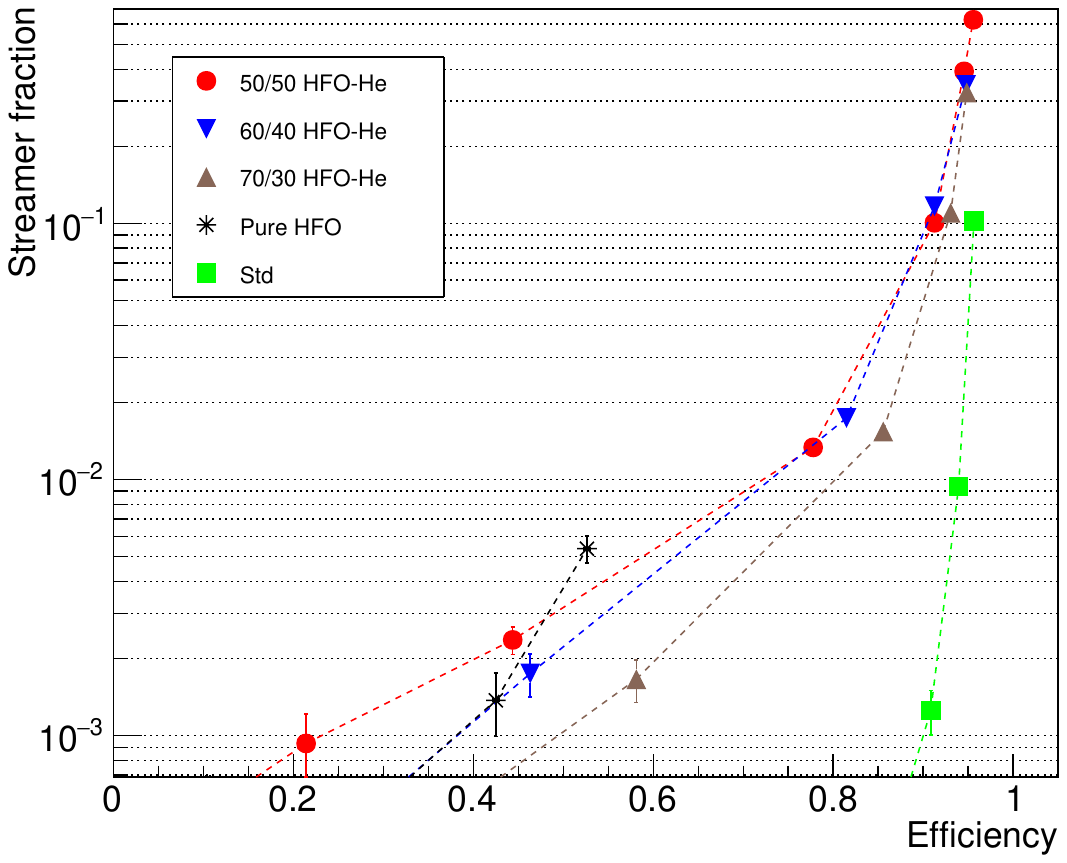}
  \caption{Streamer Vs Efficiency}
  \label{fig:ClusterStreamS4}
\end{subfigure}
\caption{ Cluster size (top) and streamer fraction (bottom), as a function of the HV (left) and efficiency (right).}
\label{fig:ClusterStream}
\end{figure}

\begin{figure}
\begin{subfigure}{.5\textwidth}
  \centering
  \includegraphics[width=1\linewidth]{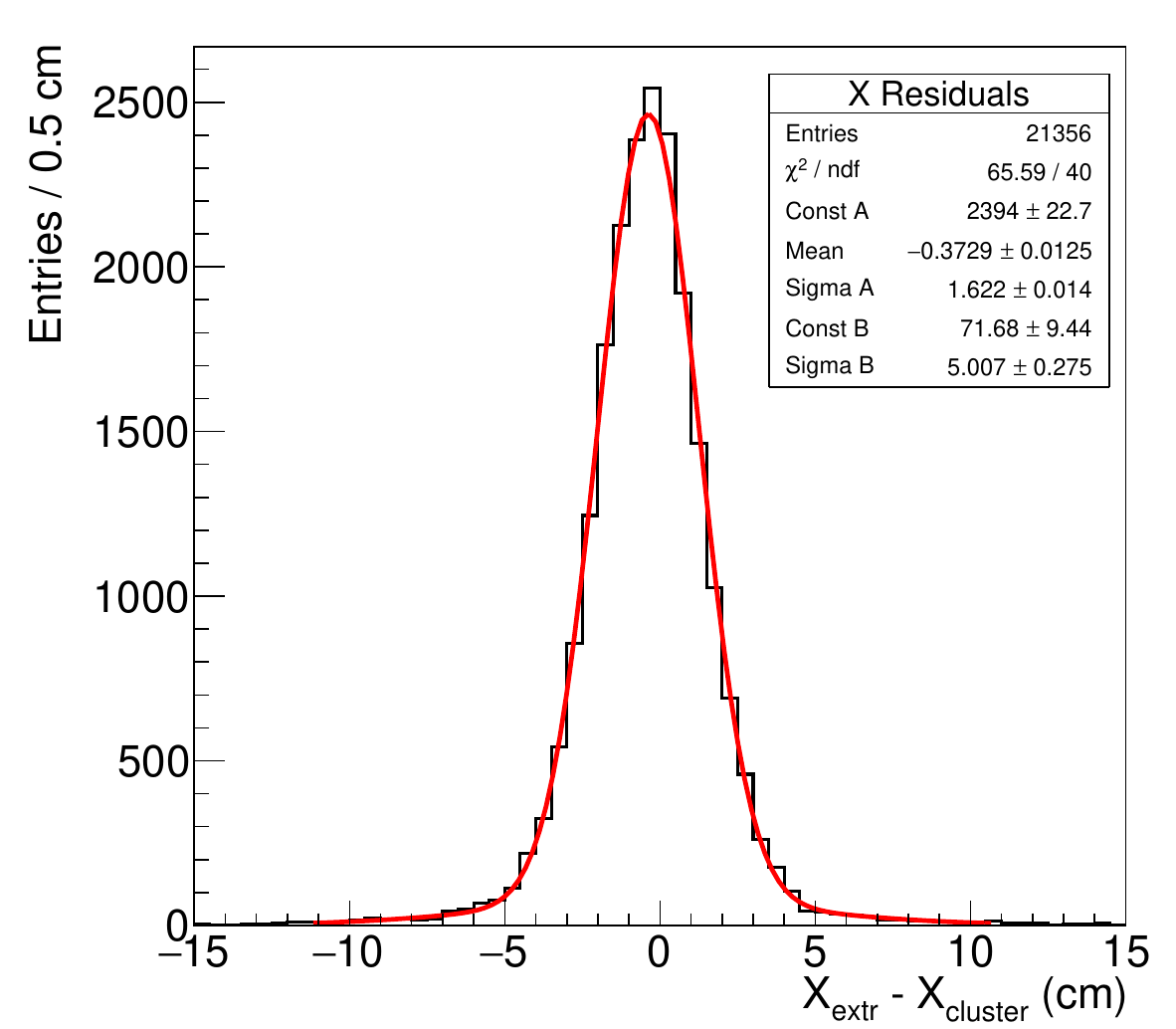}
  \caption{Std}
  \label{fig:SpatialResidualsS1}
\end{subfigure}%
\begin{subfigure}{.5\textwidth}
  \centering
  \includegraphics[width=1\linewidth]{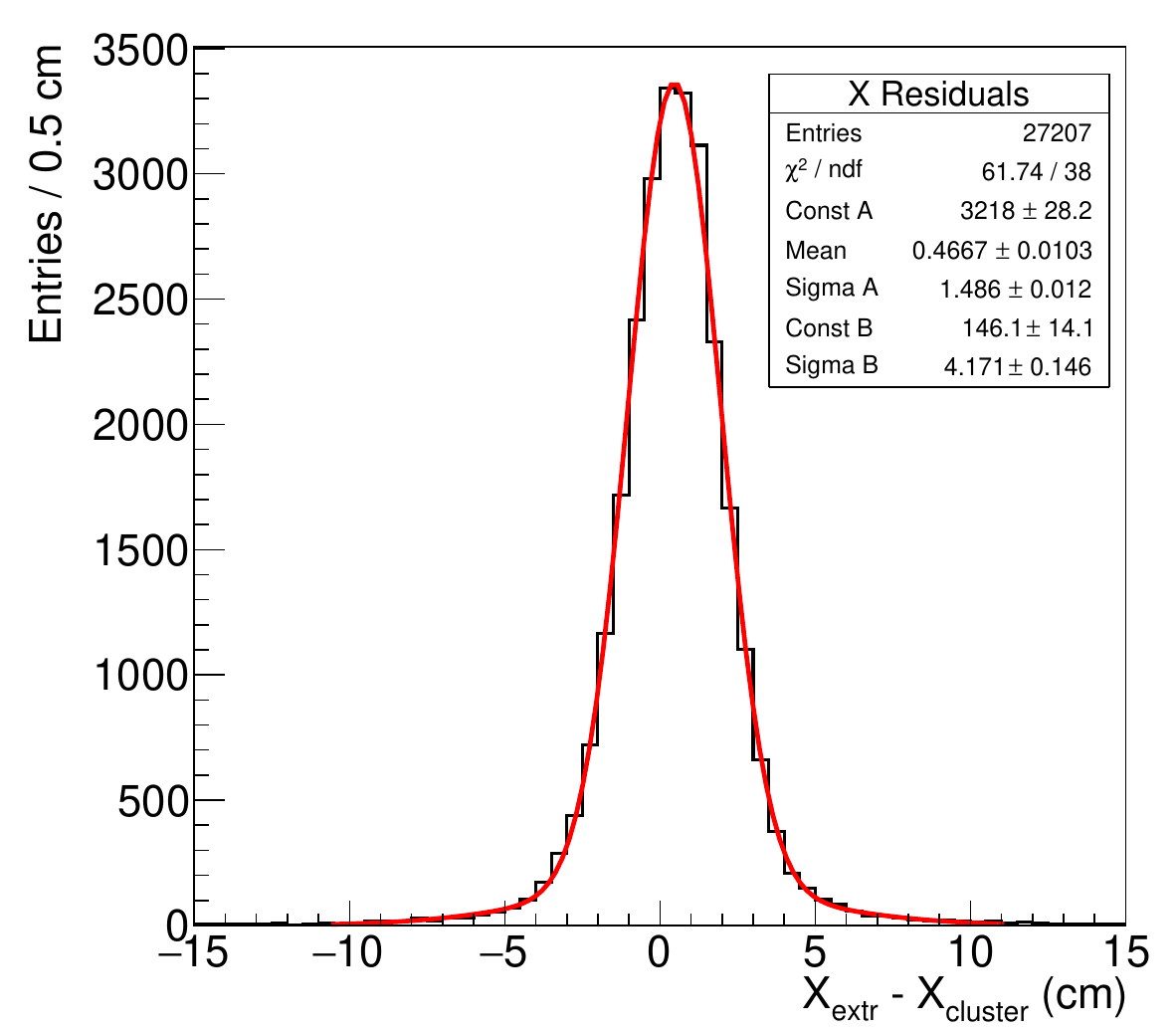}
  \caption{70/30}
  \label{fig:SpatialResidualsS2}
\end{subfigure}
\begin{subfigure}{.5\textwidth}
  \centering
  \includegraphics[width=1\linewidth]{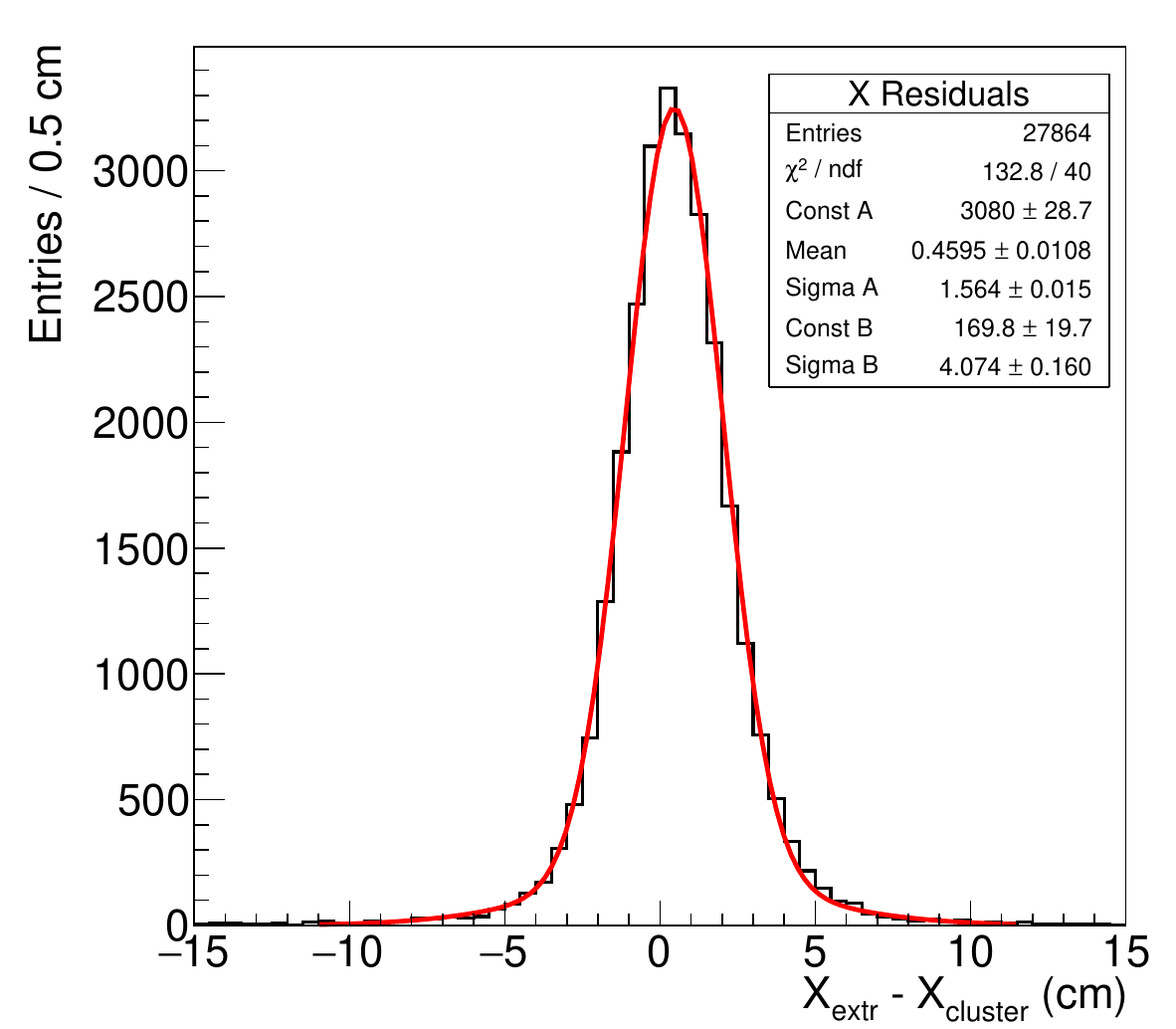}
  \caption{60/40}
  \label{fig:SpatialResidualsS3}
\end{subfigure}
\begin{subfigure}{.5\textwidth}
  \centering
  \includegraphics[width=1\linewidth]{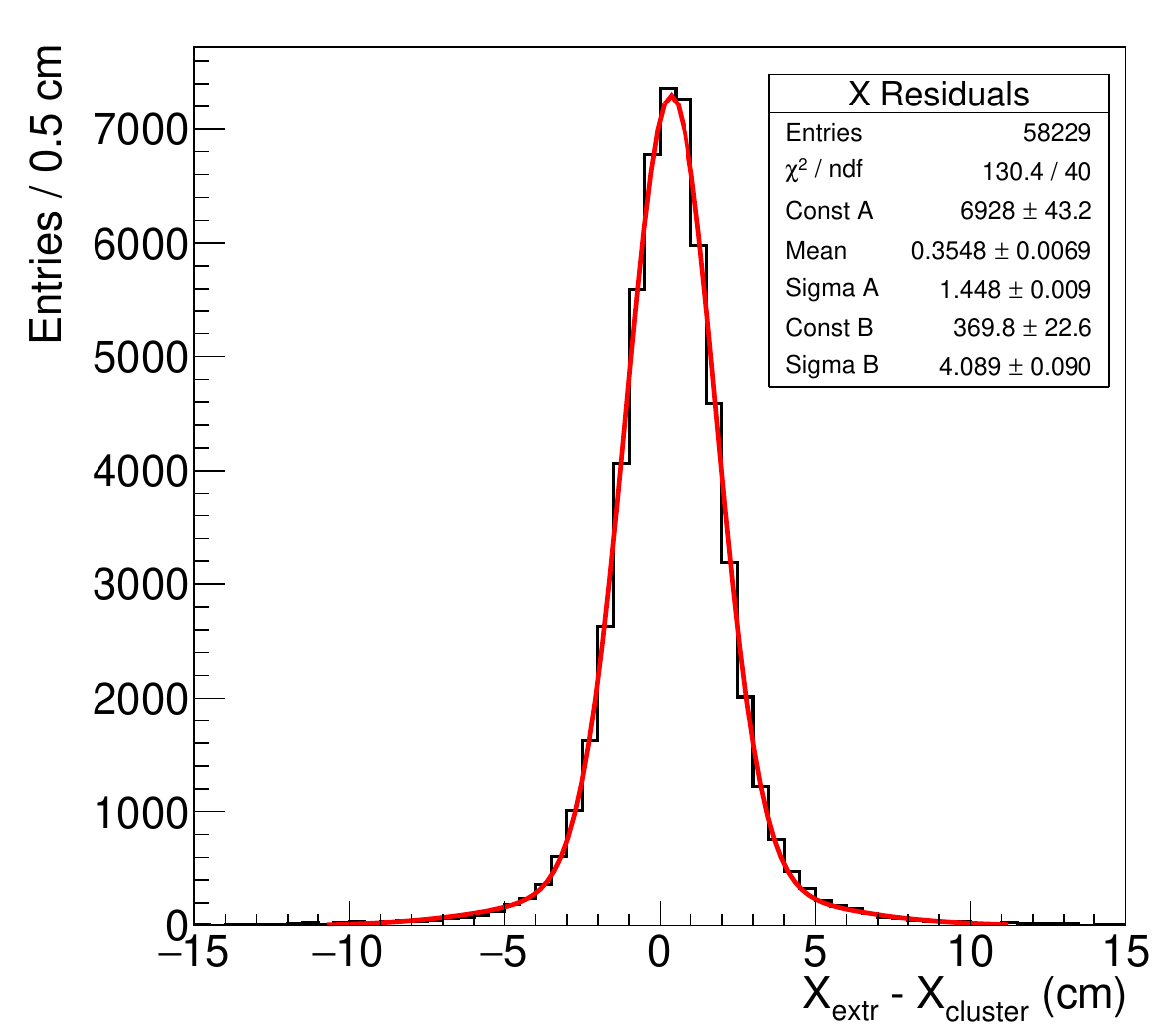}
  \caption{50/50}
  \label{fig:SpatialResidualsS4}
\end{subfigure}
\caption{Distribution of the spatial residuals in the longitudinal X coordinate for the chamber under test, using the standard mixture and 3 HFO-He mixtures}
\label{fig:SpatialResiduals}
\end{figure}

As discussed in Sec. \ref{sec:ecogas} the absence of SF$_6$ is expected to have a negative impact on the streamer probability and on the cluster size. The results reported in Fig. \ref{fig:ClusterStream} confirm this hypothesis. 
Both cluster size and streamer probability increase faster with the bias voltage than when using the standard mixture (Fig. \ref{fig:ClusterStreamS1} and \ref{fig:ClusterStreamS3}). An efficiency of 90\% can still be reached with a cluster size below $\approx3$ and a streamer fraction close to 0.1 (Fig. \ref{fig:ClusterStreamS2} and \ref{fig:ClusterStreamS4} respectively). We expect our detector to operate around this working point. While the cluster size can be easily handled by the offline clustering algorithm, the streamer fraction could pose some challenges in the reconstruction of the events, as well as for the potential aging effect of the detector. Mixtures with  percentages of He above 50\% have not been tested, since the streamer probability and cluster size are expected to exceed the allowed operation limits, and since the desired operating voltage range was already obtained.
\indent
Spatial residuals have been computed independently for the two coordinates. Residuals in the Y direction are not expected to change, being dominated by the strip quantization and estimated to be  $\approx1$ cm. This is indeed confirmed for all mixtures and voltages, except for the 50/50 mixture at higher voltage. In that condition the percentage of streamers gets above 30\% and the degradation in performance is due to a non optimal treatment of very large clusters, that is currently being addressed. Residuals in the X direction, computed using the time information as previously described, are potentially more sensitive to the change of the gas mixture. The distribution of residuals for three HFO-He mixtures and for the standard mixture are reported in Fig. \ref{fig:SpatialResiduals}. The voltages were selected  in order to obtain an efficiency of 95\%.
The double Gaussian fit is the same used in some previous  EEE study, as reported elsewhere \cite{performance}, and can successfully describe both standard and new mixtures. The standard deviation of the narrower Gaussian is in the range 1.4-1.6 cm for all mixtures, no significant differences are found.

\begin{figure}
\begin{subfigure}{.5\textwidth}
  \centering
  \includegraphics[width=1\linewidth]{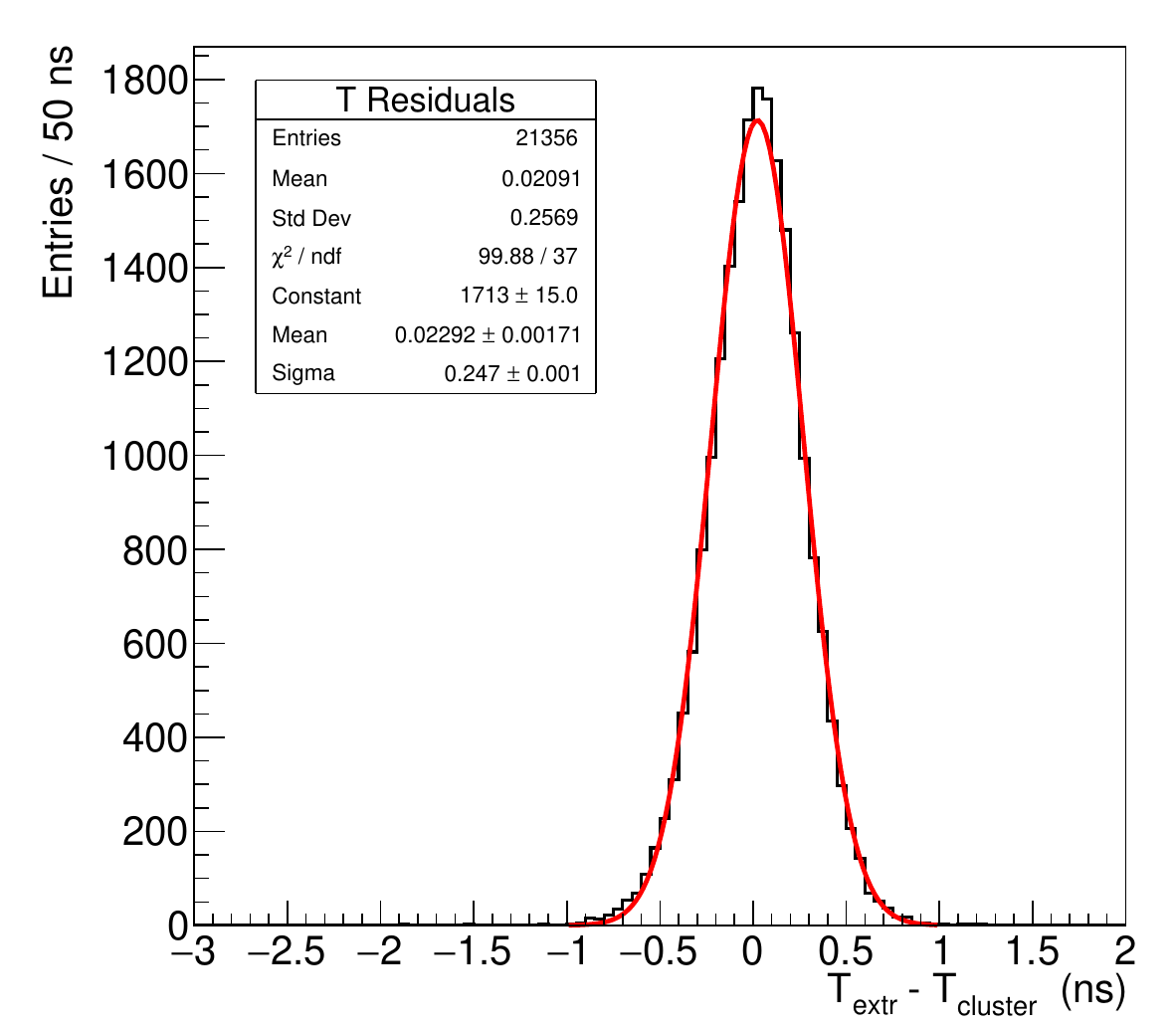}
  \caption{Std}
  \label{fig:TimingS1}
\end{subfigure}%
\begin{subfigure}{.5\textwidth}
  \centering
  \includegraphics[width=1\linewidth]{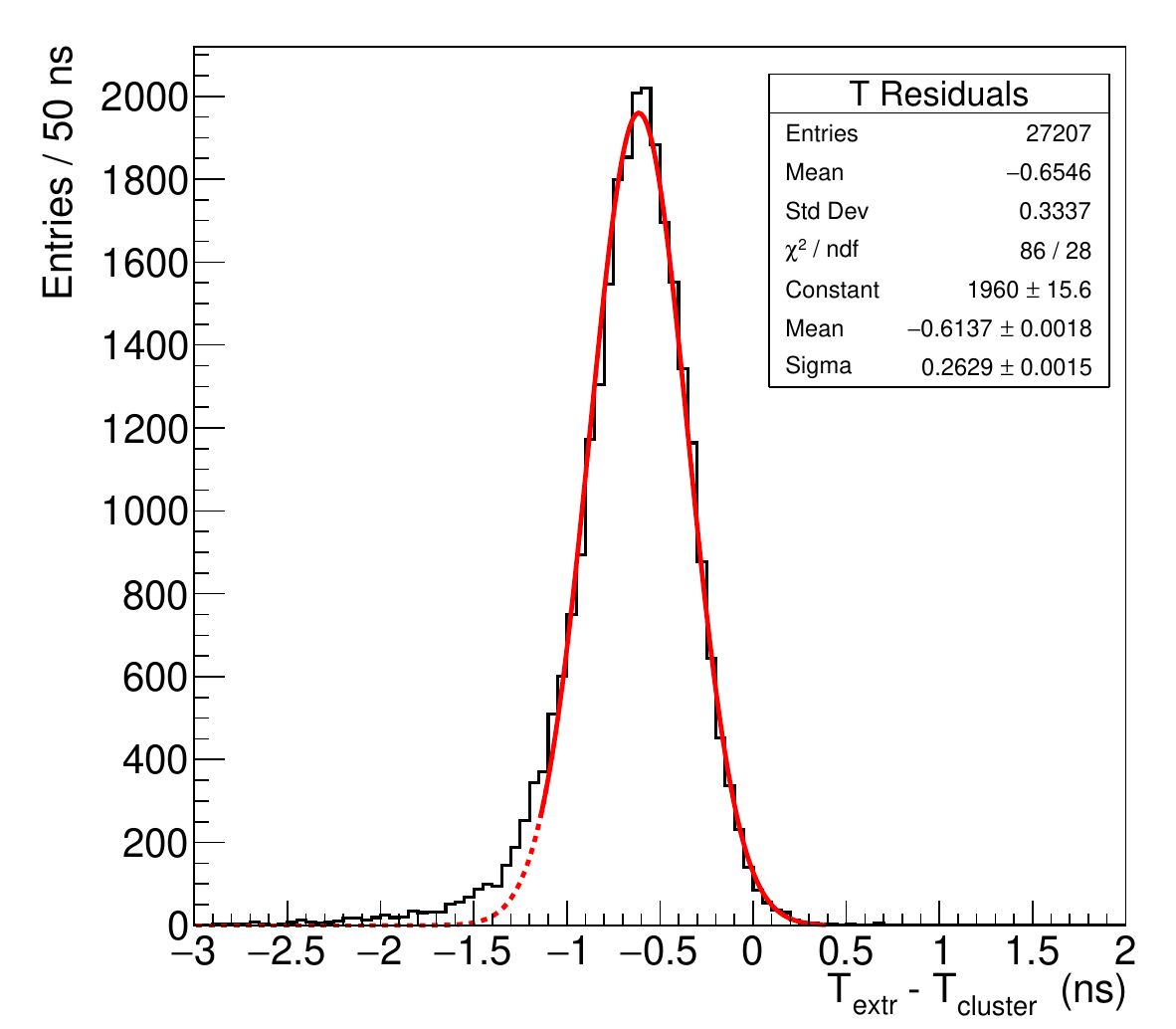}
  \caption{70/30}
  \label{fig:TimingS2}
\end{subfigure}
\begin{subfigure}{.5\textwidth}
  \centering
  \includegraphics[width=1\linewidth]{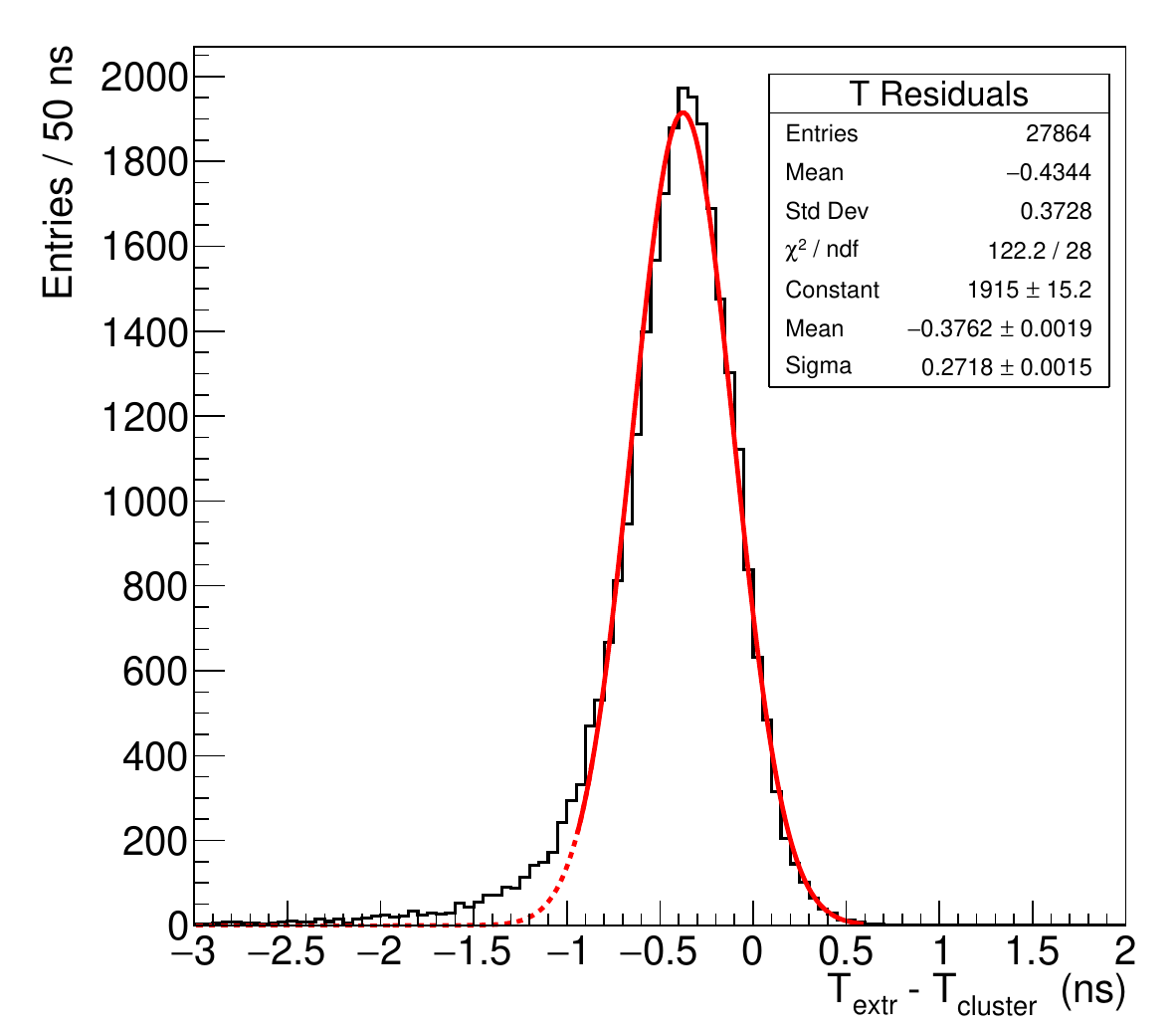}
  \caption{60/40}
  \label{fig:TimingS3}
\end{subfigure}
\begin{subfigure}{.5\textwidth}
  \centering
  \includegraphics[width=1\linewidth]{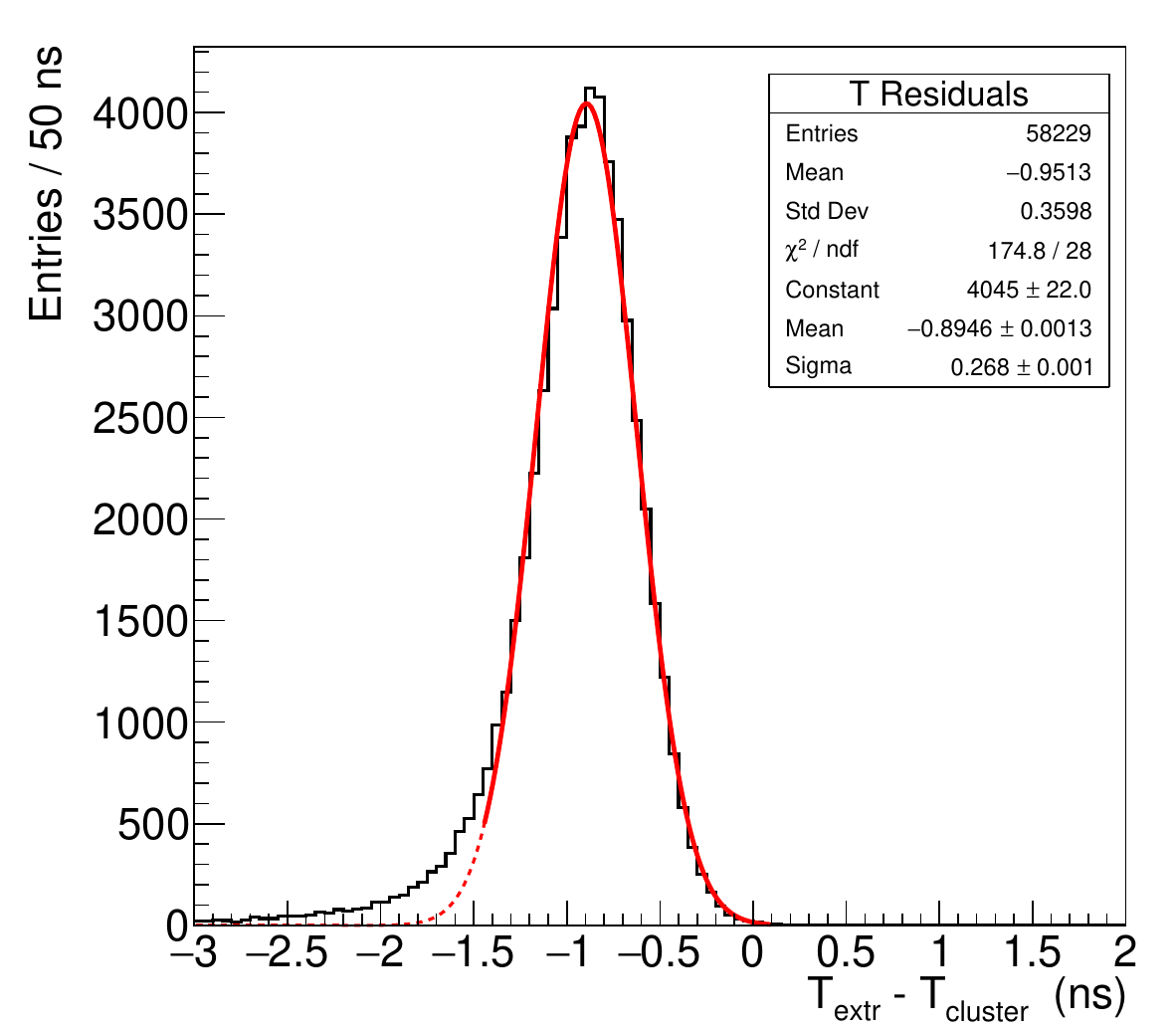}
  \caption{50/50}
  \label{fig:TimingS4}
\end{subfigure}
\caption{Distribution of the time residuals  for the chamber under test, using  the standard mixture and three different HFO-He mixtures. The dashed lines represent the extrapolation of the fits to the tail regions.}
\label{fig:Timing}
\end{figure}

Residuals have been computed, for the same data, also for the cluster time. For all mixtures a strip-by-strip time calibration has been applied. This is indeed needed to correct for possible time offsets generated by the setup (i.e. different lengths of cables or fixed offsets in TDC channels). Since such offsets are not gas dependent, the correction has been computed only once using the standard mixture and then applied to all measurements. The resulting distributions are shown in Fig. \ref{fig:Timing}.
Differently from the standard mixture, the distributions with HFO-He mixtures show a tail on the left side of the peak. Gaussian fits have been performed excluding the tails, corresponding to a fraction of outliers in the range 8-9\%.
The time residuals show a slight increase with respect to the standard mixture, suggesting a slightly lower time precision of the detector with the new mixture.
The lower time precision can be due to the different drift speed and Townsend coefficient of the new mixture, which influences the avalanche development. Another cause for the larger uncertainty can be attributed to the larger spread in signal total charge, which has been observed with the new mixtures. Further studies and offline calibrations based on TW corrections, not applied in the present analysis, can improve the detector performance, likely reducing the tails. However, no impact is expected on the absolute particle timestamp since, as discussed in Sec. \ref{sec:intro}, its uncertainty is dominated by the GPS precision. The only parameter affected will be the particle time of flight and consequently, the measurement of its speed.\\
\indent
The results show that the efficiency, the tracking performance and the capability to correlate tracks detected by different telescopes of the network are unaltered by the new mixtures, preserving the physics program of the experiment.

\section{Conclusions and outlook}\label{se:conclusions}

The EEE collaboration extensively worked on a series of tests in the last two years, with the aim to replace the standard gas mixture (98\% C$_2$H$_2$F$_4$ and 2\% SF$_6$) usually used in gaseous detectors such as MRPCs, with a new mixture with a Global Warming Potential (GWP) compatible with the new European regulation about Green Houses Gases emissions.
The EEE collaboration has demonstrated that the MRPCs composing the muon telescopes of the project can be operated with a new gas mixture of (HFO-He).
 The new mixture has the advantage of a negligible GWP and it is compliant with all project requirements.
 
The first test, performed on the REND-01 telescope, indicates that the mixture can be effectively used to replace the standard one preserving the physics program of the experiment. The spatial precision is in line with the one previously quoted \cite{performance} with the standard mixture, while the lower time precision does not impact the absolute timestamp of the particles, keeping the capability to correlate particles simultaneously detected by different telescopes of the EEE network.
Analogous tests have been conducted on the PISA-01 telescope. The data analysis is being finalized, but preliminary results are in agreement with those reported in this paper.
The new gas mixture has already been deployed to other telescopes of the network, to enhance the statistics and further confirm the results, making of EEE the first MRPC-based experiment to employ an HFO-He gas mixture for physics data taking. 

The telescopes of the EEE Project, hosted mostly in non-conventional sites as school buildings and sampling a surface of about $5\cdot10^6$ km$^2$ across the Italian territory (corresponding to about 10$^\circ$ in both latitude and longitude), are going to be back in full operation with an initial ratio of HFO-He set to 60$/$40, with further setting refinement to be done based on the collected data.

\FloatBarrier

\bibliography{Biblio}
\end{document}